\documentclass[prl,preprint]{revtex4-1}
\usepackage{graphicx}
\usepackage{appendix}
\usepackage{braket}
\usepackage{rotating}
\usepackage{simplewick}
\usepackage{amssymb}
\usepackage{amsmath}
\usepackage{txfonts}
\usepackage{bm}
\usepackage{color}
\usepackage{multirow}
\usepackage{booktabs}
\usepackage{dcolumn}

\begin{document}

%\title{A simple solution of the Kohn--Luttinger problem}
\title{General solution to the Kohn--Luttinger nonconvergence problem}
%\title{On the Kohn--Luttinger problem}

\author{So Hirata}
\email{sohirata@illinois.edu}
\affiliation{Department of Chemistry, University of Illinois at Urbana-Champaign, Urbana, Illinois 61801, USA}

\date{\today}

\begin{abstract}
A simple, but general solution is proposed for the Kohn--Luttinger problem, i.e., the nonconvergence of the finite-temperature 
many-body perturbation theory with its zero-temperature counterpart as temperature is lowered to zero under some circumstances. 
How this nonconvergence can be avoided by altering the reference wave function
is illustrated numerically by using up to the fifth order of the perturbation theory. 
\end{abstract}

\maketitle 

The Kohn--Luttinger problem \cite{kohn,luttingerward,hiratapra} refers to the eponymous authors' prediction that
the finite-temperature many-body perturbation theory for electrons \cite{matsubara,bloch,kohn,luttingerward,balian,blochbook,SANTRA}  does not always reduce to the zero-temperature counterpart
as  temperature goes to zero.
On this basis, they concluded that ``BG [Brueckner--Goldstone perturbation] series
is therefore in general not correct'' \cite{kohn}.
In this light, contemporary textbooks contain
a qualification about the validity of the very finite-temperature perturbation theory they teach. 
March, Young, and Sampanthar \cite{march} wrote, ``a note of caution is called for whenever we attempt to calculate zero-temperature properties
from an expression for the same quantities at non-zero temperature $T$ by taking the limit $T \to 0$. The physics is not necessarily 
the same in both cases.'' Thouless \cite{thouless1972quantum} was more blunt, writing, ``In any case it serves as a warning against taking the results of 
perturbation theory too seriously.'' For over 60 years since the Kohn--Luttinger paper \cite{kohn}, however,
it has been unclear whether the predicted inconsistency 
actually exists \cite{SANTRA} and if it does, what causes it.

Last year, having established \cite{JhaHirata,HirataJha,HirataJha2,HirataJCP}  the finite-temperature perturbation theory for the grand potential ($\Omega$), 
chemical potential ($\mu$), and internal energy ($U$),
we showed \cite{hiratapra} that the first- and second-order perturbation corrections 
to $U$ indeed disagree with the corresponding corrections to the ground-state energy as $T \to 0$ when the reference wave function 
 differs qualitatively from the exact one. 
In particular, when the degeneracy of the reference wave function is partially or fully lifted at the first-order degenerate Hirschfelder--Certain perturbation theory \cite{Hirschfelder}, 
the second-order corrections to $\Omega$ and $U$ are divergent \cite{hiratapra}. 
The cause of this nonconvergence or divergence is traced to the fact that the definitions of $\Omega$ and $U$  
are nonanalytic at $T = 0$ and cannot be expanded in a converging power series. % of perturbation strength $\lambda$ when the ground-state energy touches or crosses an excited-state energy as $\lambda = 1 \to 0$. 
Therefore, the Kohn--Luttinger problem does exist, and may be defined more broadly:\ The finite-temperature 
many-body perturbation theory has zero radius of convergence at $T = 0$ for a qualitatively wrong reference \cite{hiratapra}. 

In this Letter, we first show that the divergence due to a degenerate reference
occurs at higher order of the finite-temperature perturbation theory, and verify it numerically using 
its general-order algorithm \cite{HirataJCP}. 
We then demonstrate how this divergence can be avoided by simply changing the reference to a nondegenerate one.
Since the singlet and triplet instability theorems \cite{YamadaHirata} guarantee  that for any given degenerate Hartree--Fock (HF) wave function there is always
a lower-lying nondegenerate wave function, the proposed solution is general.
%, and possibly to a homogeneous electron gas as per the Overhauser theorem \cite{Overhauser1,Overhauser2}. 
We shall consider the perturbation corrections to $U$ in the grand canonical ensemble \cite{HirataJCP}, but what follows applies equally to $\Omega$ and to the canonical ensemble \cite{JhaHirata_canonical}.

The general-order algorithm implements the recursion relation \cite{HirataJCP} that defines the $n$th-order correction to the grand potential $\Omega^{(n)}$ 
in terms of lower-order corrections, which reads
\begin{eqnarray}
\Omega^{(n)} &=& \langle D^{(n)} \rangle + \frac{(-\beta)}{2!} \sum_{i=1}^{n-1} \left( \langle D^{(i)}D^{(n-i)} \rangle - \Omega^{(i)}\Omega^{(n-i)} \right) 
\nonumber\\&& 
+ \frac{(-\beta)^2}{3!} \sum_{i=1}^{n-2}\sum_{j=1}^{n-i-1} \left( \langle D^{(i)}D^{(j)} D^{(n-i-j)} \rangle - \Omega^{(i)}\Omega^{(j)} \Omega^{(n-i-j)}\right) \nonumber\\&&
+ \dots + \frac{(-\beta)^{n-1}}{n!}  \left\{ \langle (D^{(1)})^n\rangle - (\Omega^{(1)})^n  \right\},  \label{recursionOmega}
\end{eqnarray}
where $\beta = 1/k_\text{B}T$ and $\langle \dots \rangle$ denotes the zeroth-order thermal average, i.e.,
\begin{eqnarray}
\langle X \rangle =  \frac{\sum_IX_I e^{-D_I^{(0)}}}{\sum_I e^{-D_I^{(0)}}}  ,
\end{eqnarray}
where $I$ runs over all states, $D_I^{(n)} = E_I^{(n)} - \mu^{(n)} N_I$, $E_I^{(n)}$ is the $n$th-order Hirschfelder--Certain degenerate perturbation correction \cite{Hirschfelder} to the $I$th-state energy,
$\mu^{(n)}$ is the $n$th-order perturbation correction to the chemical potential, and $N_I$ is the number of electrons in the $I$th state.
Starting with $\Omega^{(0)}$ and $\mu^{(0)}$ furnished by the Fermi--Dirac theory, arbitrarily high orders of $\Omega^{(n)}$ can be generated.  
It is important to base our theory on the Hirschfelder--Certain degenerate perturbation theory \cite{Hirschfelder} because the latter 
is the proper Rayleigh--Schr\"{o}dinger perturbation theory for degenerate and nondegenerate references.
While unimportant in this study, in deriving reduced analytical formulas, $E_I^{(n)}$ can be replaced by the corresponding diagonal element of the effective Hamiltonian matrix of 
the degenerate perturbation theory \cite{HirataJha,HirataJha2,HirataJCP}, which can be expressed in a closed form by the Slater--Condon rules.

The recursion for $U^{(n)}$ is given \cite{HirataJCP} by
\begin{eqnarray}
U^{(n)}  &=&
 \langle  E^{(n)} \rangle + {(-\beta)}\sum_{i=1}^{n} \langle D^{(i)}D^{(n-i)} \rangle 
 \nonumber\\&&
- (-\beta) \sum_{i=1}^{n} \Omega^{(i)} (U^{(n-i)}-\mu^{(n-i)}\bar{N})  
 \nonumber\\&&
+ \frac{(-\beta)^2}{2!} \sum_{i=1}^{n-1}\sum_{j=1}^{n-i} \langle D^{(i)}D^{(j)} D^{(n-i-j)} \rangle 
\nonumber\\&&
- \frac{(-\beta)^2}{2!} \sum_{i=1}^{n-1}\sum_{j=1}^{n-i} \Omega^{(i)} \Omega^{(j)} (U^{(n-i-j)} -\mu^{(n-i-j)}\bar{N}) 
\nonumber\\&&
%+ \frac{(-\beta)^3}{3!} \sum_{i=1}^{n-2}\sum_{j=1}^{n-i-1}\sum_{k=1}^{n-i-j} \langle D^{(i)}D^{(j)} D^{(k)}    D^{(n-i-j-k)} \rangle 
%- \frac{(-\beta)^3}{3!} \sum_{i=1}^{n-2}\sum_{j=1}^{n-i-1}\sum_{k=1}^{n-i-j} \Omega^{(i)}\Omega^{(j)} \Omega^{(k)}    (U^{(n-i-j-k)}-\mu^{(n-i-j-k)}\bar{N}) \rangle 
%\nonumber\\&&
+ \dots 
+ \frac{(-\beta)^n}{n!}  \langle (D^{(1)})^n D^{(0)} \rangle 
\nonumber\\&&
- \frac{(-\beta)^n}{n!}   (\Omega^{(1)})^n (U^{(0)}-\mu^{(0)}\bar{N}), \label{recursionU}
\end{eqnarray}
where $\bar{N}$ is the average number of electrons that keeps the system electrically neutral.
Note that $\Omega^{(1)} = \langle D^{(1)} \rangle$ and $U^{(0)}-\mu^{(0)}\bar{N} = \langle D^{(0)} \rangle$. 
Similar recursions exist for $\mu^{(n)}$ and $S^{(n)}$ \cite{HirataJCP}. 

\begin{figure}
\includegraphics[scale=1.3]{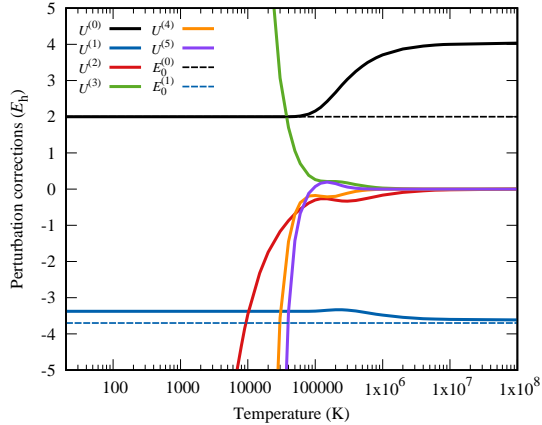}
\caption{$U^{(n)}$ $(0 \leq n \leq 5)$ as a function of temperature with a degenerate Hartree--Fock reference of the square-planar H$_4$. \label{fig:1}}
\end{figure}

Figure \ref{fig:1} shows $U^{(n)}$ ($0 \leq n \leq 5$) as a function of $T$ for an ideal gas of square-planar H$_4$ (0.8 \AA\ in the STO-3G basis set) 
with a degenerate HF reference. The ideal gas is an ensemble of an infinite number of rigid and nonrotating H$_4$ molecules that do not interact with
one another except to export or import electrons. Our previous paper \cite{hiratapra} included the same data for $0 \leq n \leq 2$ using sum-over-orbital formulas, 
and this study extends the calculation to the fifth-order 
perturbation theory with the general-order algorithm \cite{HirataJCP}. 
The dashed lines indicate the correct zero-temperature limits of $U^{(0)}$ and $U^{(1)}$, which are determined by the zero-temperature  Hirschfelder--Certain
degenerate perturbation theory \cite{Hirschfelder} and are denoted by $E^{(0)}$ and $E^{(1)}$, respectively. 

As shown before \cite{hiratapra}, $U^{(0)}$, i.e., the Fermi--Dirac theory converges correctly at $E^{(0)}$ as $T \to 0$, whereas 
$U^{(1)}$ approaches a finite, but incorrect zero-temperature limit, which differs 
from the correct limit of $E^{(1)}$. Worse yet, $U^{(2)}$ is divergent, while the correct limit $E^{(2)}$ is finite. 
In this study, it is  shown that $U^{(3)}$ through $U^{(5)}$ as well as all higher-order corrections 
diverge and thus have a wrong zero-temperature limit. 

From an analytical viewpoint, $\Omega^{(n)}$ and $U^{(n)}$ contain
$\beta\,\text{cov}(D^{(i)},D^{(n-i)})$ (where `cov' stands for the covariance within a degenerate subspace) \cite{hiratapra,HirataJCP}, 
which is responsible for the divergence at $\beta = \infty$ in the event that both $D^{(i)}$ and $D^{(n-i)}$ 
have a distribution within the degenerate subspace of the reference state. 
When the degeneracy of the reference is partially or fully lifted for the first time at the $n$th order of the Hirschfelder--Certain degenerate perturbation
theory \cite{Hirschfelder}, giving $D^{(n)}$ a distribution, $\Omega^{(2n)}$ and $U^{(2n)}$ become divergent at $T = 0$.  When the reference state is nondegenerate, $D_0^{(i)}$ and $D_0^{(n-i)}$ (subscript 0 standing for the reference)
are just single numbers and cannot have a distribution, making all terms carrying a factor of $\beta$ vanish as $T \to 0$. Then, $\Omega^{(n)}$ or $U^{(n)}$ are no longer divergent. 

Since one is normally interested in $\Omega$ and $U$ at lower temperatures
first, the theory is rather useless when the reference is degenerate. It may be said that the pathological behavior of the perturbation theory 
is amplified by finite temperature, considering the fact that 
the zero-temperature Hirschfelder--Certain  degenerate perturbation theory \cite{Hirschfelder} is rapidly convergent for the same degenerate reference. 

\begin{figure}
\includegraphics[scale=1.3]{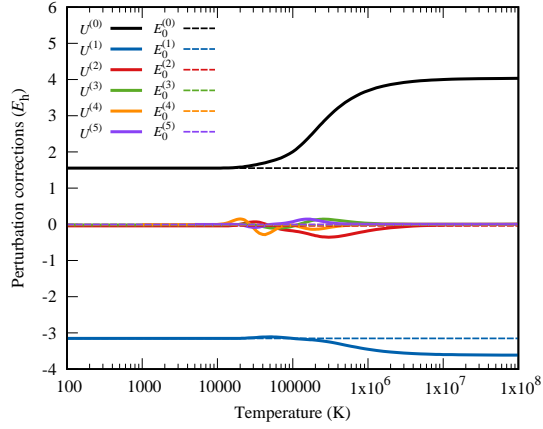}
\caption{$U^{(n)}$ $(0 \leq n \leq 5)$ as a function of temperature with a nondegenerate Hartree--Fock reference of the square-planar H$_4$. \label{fig:2}}
\end{figure}

\begin{figure}
\includegraphics[scale=1.3]{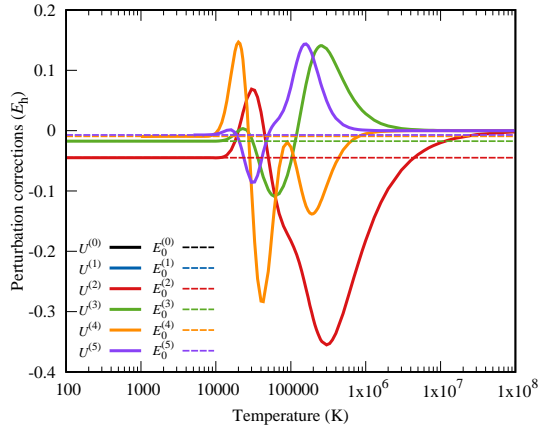}
\caption{The same as Fig.\ \ref{fig:2} with a different range. \label{fig:3}}
\end{figure}

One can, however, easily avoid this nonconvergence or divergence by altering the reference. With no symmetry restriction, the HF calculation naturally
converges to a symmetry-broken nondegenerate wave function with an energy of $-1.6017\,E_\text{h}$. The latter 
is much lower than the  energy of $-1.3791\,E_\text{h}$ of the degenerate reference used in Fig.\ \ref{fig:1}. 
See Ref.\ \cite{hiratapra} for more details about the degenerate reference.
Starting from the nondegenerate reference, the finite-temperature perturbation theory for $U^{(n)}$ converges at the 
correct zero-temperature limit of $E^{(n)}$ provided by the nondegenerate (M{\o}ller--Plesset) perturbation theory \cite{moller} for all $0 \leq n \leq 5$ considered. 
This is shown in Figs.\ \ref{fig:2} and \ref{fig:3}. (The oscillations of $U^{(n)}$ seen in the middle of the graph are due to the temperature reaching 
the lowest excitation energy for the first time, causing the perturbation series to converge more slowly.)

It may be emphasized that this simple solution to the Kohn--Luttinger problem is not specific to the square-planar H$_4$, but is generally
applicable to any finite, degenerate system. This is because the singlet and triplet instability theorems \cite{YamadaHirata} state that for any given HF solution
with degenerate (or form-degenerate) 
highest occupied and lowest unoccupied molecular orbitals, there exists a lower-lying, symmetry-broken solution with the degeneracy lifted. 
The degenerate solution is only metastable, and it is actually easier to find a nondegenerate HF solution with a lower energy. 
If ``strong correlation'' is characterized by the extent to which a perturbation theory fails to converge, the Kohn--Luttinger problem
may be viewed as a particularly severe case of strong-correlation problems. % because the zero-temperature degenerate perturbation theory \cite{Hirschfelder} is rapidly convergent  for the same degenerate reference. 
Our general solution, however, suggests that this  %, like many others if not all, 
is a manufactured problem caused by a metastable degenerate reference, and is best addressed at the level of mean-field theories.
For a homogeneous electron gas, in which Kohn and Luttinger \cite{kohn} originally considered the potential issue of nonconvergence, 
the Overhauser theorems \cite{Overhauser1, Overhauser2} also guarantee a lower-energy nondegenerate HF solution.

\begin{figure}
\includegraphics[scale=1.3]{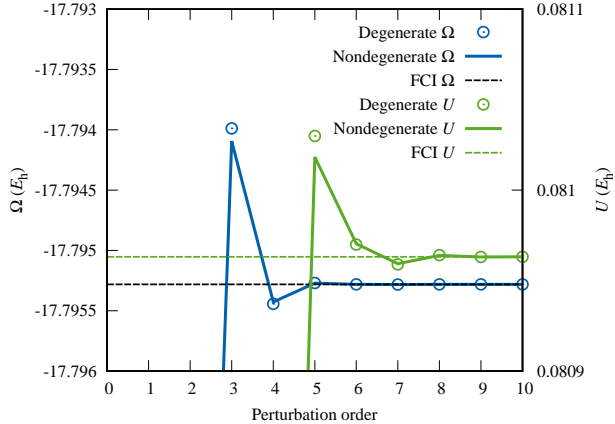}
\caption{Convergence of the perturbation approximations to $\Omega$ and $U$ in comparison with the thermal full-configuration-interaction values 
of the square-planar H$_4$ at $T = 10^6\,\text{K}$. \label{fig:4}}
\end{figure}

Finally, Fig.\ \ref{fig:4} verifies the convergence of the perturbation series of $\Omega$ and $U$ at the exact limits set by
the thermal full configuration interaction \cite{Kou} at a given temperature. 
The temperature of $10^6\,\text{K}$ was chosen only because the finite-temperature perturbation theory happens to be nondivergent there.
The latter method evaluates the grand partition function with the exact energies 
of all states with any number of electrons (made available by the zero-temperature full-configuration-interaction method) and derives thermodynamic functions from it \cite{Kou}. 
That the series reach the same numerically exact values
at the tenth order either from the degenerate or nondegenerate reference underscores the correctness of the finite-temperature perturbation theory 
introduced by us \cite{HirataJha,HirataJha2,HirataJCP}, not overlooking any diagrammatic contribution such as anomalous and renormalization diagrams.

It is important to distinguish two types of convergence discussed in this Letter: The convergence of the finite-temperature perturbation theory towards
the zero-temperature perturbation theory as $T \to 0$, which is the focus of this study; the convergence of the finite-temperature perturbation theory 
towards the exact (thermal-full-configuration-interaction) limit at a given temperature, which is illustrated in Fig.\ \ref{fig:4}. 
The former is not expected for a qualitatively wrong reference \cite{hiratapra}, but restored by changing the reference. 
This has been argued mathematically on the basis of the recursion, and verified numerically for up to the fifth order. 
The latter convergence towards the exact limit 
is guaranteed (unless the perturbation series diverges \cite{olsen_mp}) because the finite-temperature perturbation theory \cite{HirataJha,HirataJha2,HirataJCP} is correct.

\acknowledgments

%I thank Dr.\ Punit K. Jha for many insightful discussions. 
This work was supported by the U.S. Department of Energy, Office of Science, Office of Basic Energy Sciences under Grant No.\ DE-SC0006028
and also by the Center for Scalable, Predictive methods for Excitation and Correlated phenomena (SPEC), which is funded by 
the U.S. Department of Energy, Office of Science, Office of Basic Energy Sciences, 
Chemical Sciences, Geosciences, and Biosciences Division, as a part of the Computational Chemical Sciences Program.

\bibliography{library.bib}

\end{document}